\documentclass{cernrep}

\RequirePackage{lineno}

\usepackage[detect-all]{siunitx}  
\usepackage{enumitem}
\usepackage{hyperref}
\usepackage{wrapfig}
\begin{document}
\setcounter{page}{0}
\modulolinenumbers[1]

\begin{center}
{\LARGE Feasibility Study for an EDM Storage Ring}
\break 
\vskip0.2in

Contact:\; Hans Str\"oher\break h.stroeher@fz-juelich.de

\end{center}


\begin{center}
 for the Charged Particle Electric Dipole Moment Collaboration\footnote{\htmladdnormallink{http://pbc.web.cern.ch/edm/edm-default.htm}{http://pbc.web.cern.ch/edm/edm-default.htm}}\\
 \end{center}

\centerline{\includegraphics[width=0.4\textwidth]{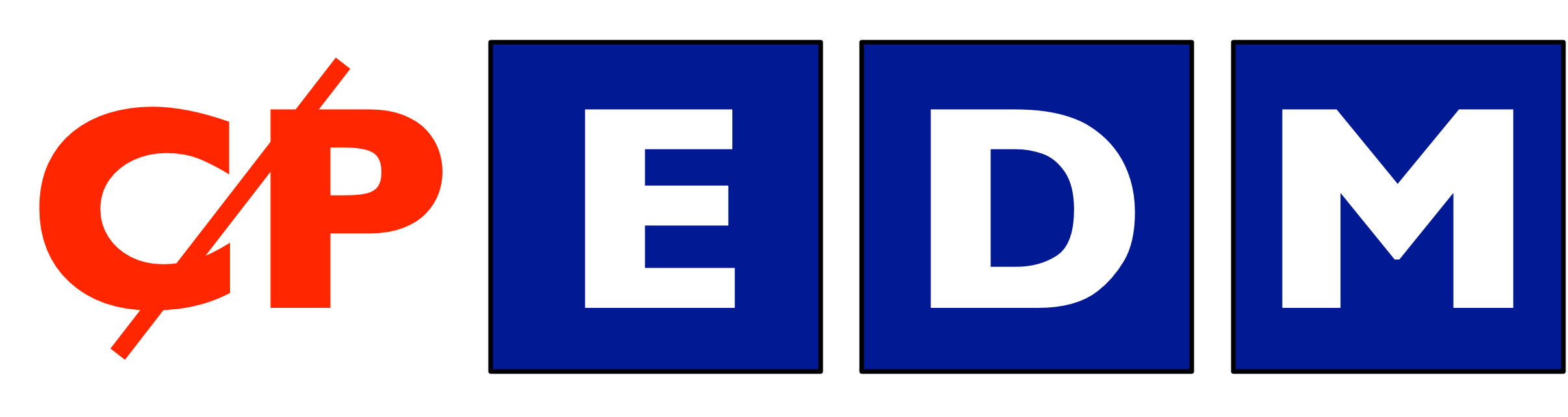}}

\vskip1.3in
\begin{center}
\begin{minipage}{4.3in}
\noindent {\bf Abstract}
\vskip0.2in

\noindent This project exploits charged particles confined as a storage ring beam
(proton, deuteron, possibly $^3$He) to search for an intrinsic electric dipole moment
(EDM, $\vec d$) aligned along the particle spin axis. Statistical sensitivities can approach
$10^{-29}$~e$\cdot$cm. The challenge will be to reduce 
systematic errors to similar levels. The ring will be adjusted to
preserve the spin polarization, initially parallel to the particle velocity,
for times in excess of 15 minutes. Large radial electric fields, acting
through the EDM, will rotate the polarization ($\vec d \times\vec E$).
The slow rise in the vertical polarization component, detected through scattering from
a target, signals the EDM. 
The project strategy is
outlined. It foresees a step-wise plan, starting with ongoing COSY
activities that demonstrate technical feasibility. 
Achievements to date include reduced polarization measurement
errors, long horizontal-plane polarization lifetimes, and control of
the polarization direction through feedback from the scattering
measurements. The project continues with a proof-of-capability measurement
(precursor experiment; first direct deuteron EDM measurement), an intermediate prototype ring 
(proof-of-principle; demonstrator for
key technologies), and finally the 
high precision electric-field storage ring.
\end{minipage}
\end{center}

\thispagestyle{empty} 

%
\label{Chap:ExecSum}





\section{Science context and objectives}


Symmetry considerations and symmetry-breaking patterns have played an important role in the development
of physics in the last 100 years.
Experimental tests of discrete
    symmetries ({\it e.g.}\ parity P, charge-conjugation C, their product CP,
time-reversal invariance T, the product CPT,
baryon- and/or lepton number) have been essential for the development of
the Standard Model (SM) of particle physics.

Subatomic particles with nonzero spin (regardless whether of elementary or composite nature)
can only support a nonzero permanent electric dipole moment (EDM)
if both  time-reversal (T) and parity (P) symmetries  are violated explicitly while
the charge symmetry (C) can be maintained (see {\it e.g.}~\cite{Eng}).  Assuming the conservation of
the combined CPT symmetry, T-violation also implies CP-violation.
The  CP-violation generated by the Kobayashi-Maskawa  (KM) mechanism of weak interactions
contributes a very small EDM that is several orders of magnitude
below current experimental limits. However, many models beyond the Standard Model
predict EDM values near the current experimental limits. 
Finding a non-zero EDM value of any subatomic particle would be a signal that there exists
a new source of CP violation, either  induced by the
strong CP violation via the $\theta_{QCD}$ angle
or by genuine physics beyond the SM (BSM).
In fact, the best upper limit on 
$\theta_{QCD}$ follows from the experimental bound on the EDM of the neutron.
CP violation beyond the SM is also essential for explaining the mystery of the observed baryon-antibaryon asymmetry of our universe, 
one of the outstanding problems in contemporary elementary particle physics and cosmology. A measurement of a single EDM will not be sufficient
to establish the sources of any new CP-violation. 
Complementary observations of EDMs in multiple systems will thus prove essential. Up to now measurements have focused on neutral systems (neutron, atoms, molecules). We propose to use a storage ring to measure the EDM of charged particles.

The storage ring method would provide a direct measurement of the EDM of a charged particle comparable to or better than present investigations on ultra-cold neutrons. The neutron investigations measure the precession frequency jumps in traps containing magnetic and electric fields as the sign of the electric field is changed. These experiments are now approaching sensitivities of $10^{-26}$~e$\cdot$cm~\cite{Pen} and promise improvements of another order of magnitude within the next decade. Because proton beams trap significantly more particles, statistical sensitivities may reach the order of $10^{-29}$~e$\cdot$cm~\cite{Ana} with a new, all-electric, high-precision storage ring. Indirect determinations for the proton 
produce model-dependent EDM limits near $2 \times 10^{-25}$~e$\cdot$cm using $^{199}$Hg~\cite{And}. Thus storage rings could take the lead as the most sensitive method for the discovery of an EDM.

It should be noted that the rotating spin-polarized beam used in the EDM search is also sensitive to the presence of an oscillating EDM resulting from axions or axion-like fields, which correspond to the dark-matter candidates of a pseudo-scalar nature. These may be detected through a time series analysis of EDM search data or by scanning the beam's spin-rotation frequency in search of a resonance with an axion-like mass in the
range from $\mu$eV down to $10^{-24}$~eV~\cite{Cha,Abe}.

\section{Methodology}

The electric dipole must be aligned with the particle spin since it provides the only axis in its rest frame. The EDM signal is based on the rotation of the dipole in the presence of an external electric field that is perpendicular to the particle spin. The particles are formed into a spin-polarized beam. Measurements are made on the beam as it circulates in the ring, confined by the ring electro-magnetic fields that always generate an electric field in the particle frame pointing to the center of the ring.

The spin motion of particles in a circular accelerator or storage ring  is described by the Thomas-BMT equation and its extension for the EDM~\cite{Fuk}: 
\begin{eqnarray}
\dfrac{d \vec{S}}{dt} &=& (\vec{\Omega}_{MDM} + \vec{\Omega}_{EDM}) \times \vec{S},  \label{eq2}\\  
\vec{\Omega}_{MDM} &=& -\dfrac{q}{m} ~ \left[G \vec{B} -\dfrac{\gamma G}{\gamma+1} \vec{\beta} \left(\vec{\beta} \cdot \vec{B} \right) - \left(G-\dfrac{1}{\gamma^2-1} \right) \dfrac{\vec{\beta} \times \vec{E}}{c}\right], \nonumber\\
\vec{\Omega}_{EDM} &=& -\dfrac{\eta q}{2 m c} \left[\vec{E} - \dfrac{\gamma}{\gamma+1} \vec{\beta} \left(\vec{\beta} \cdot \vec{E}\right)+ c \vec{\beta} \times \vec{B} \right].
\nonumber
\end{eqnarray}
$\vec{S}$ in this equation denotes the spin vector in the particle rest frame, $t$ the time
in the laboratory system, $\beta$ and $\gamma$ the relativistic Lorentz factors, and $\vec{B}$ and $\vec{E}$ the magnetic and electric fields, respectively.
The magnetic anomaly $G$  and the electric dipole $\eta$ are dimensionless and introduced via the magnetic dipole moment $\vec{\mu}$ and electric dipole moment $\vec{d}$, which are both pointing in the same direction and are proportional to the particle's spin $\vec{S}$:
\begin{equation}
\begin{array}{c} 
\vec{\mu} = g \dfrac{q \hbar}{2 m} \vec{S} = (1+G) \dfrac{q \hbar}{m} \vec{S},\;\;\;\;\; \vec{d} = \eta \dfrac{q \hbar}{2 m c} \vec{S},
\label{eq1}
\end{array}
\end{equation}
where $q$ and $m$ are the charge and the mass of the particle, respectively. 
 
The effect of the torque is shown in Fig.~1 where {\bf v} is the particle velocity along the orbit, {\bf B} and {\bf E} are possible external fields (acting on a positively charged particle), and the spin axis is given by the purple arrow that rotates upward in a plane perpendicular to {\bf E}. If the initial condition begins with the spin parallel to the velocity, then the rotation caused by the EDM will make the vertical component of the beam polarization change. This becomes the signal observed by a polarimeter located in the ring. This device allows beam particles to scatter from nuclei in a fixed bulk material target (black). The difference in the scattering rate between the left and right directions (into the blue detectors) is sensitive to the vertical polarization component of the beam. Continuous monitoring will show a change in the relative left-right rate difference during the time of the beam storage if a measurable EDM is present.

\begin{figure}[htb]
\centering
\includegraphics[scale=0.7]{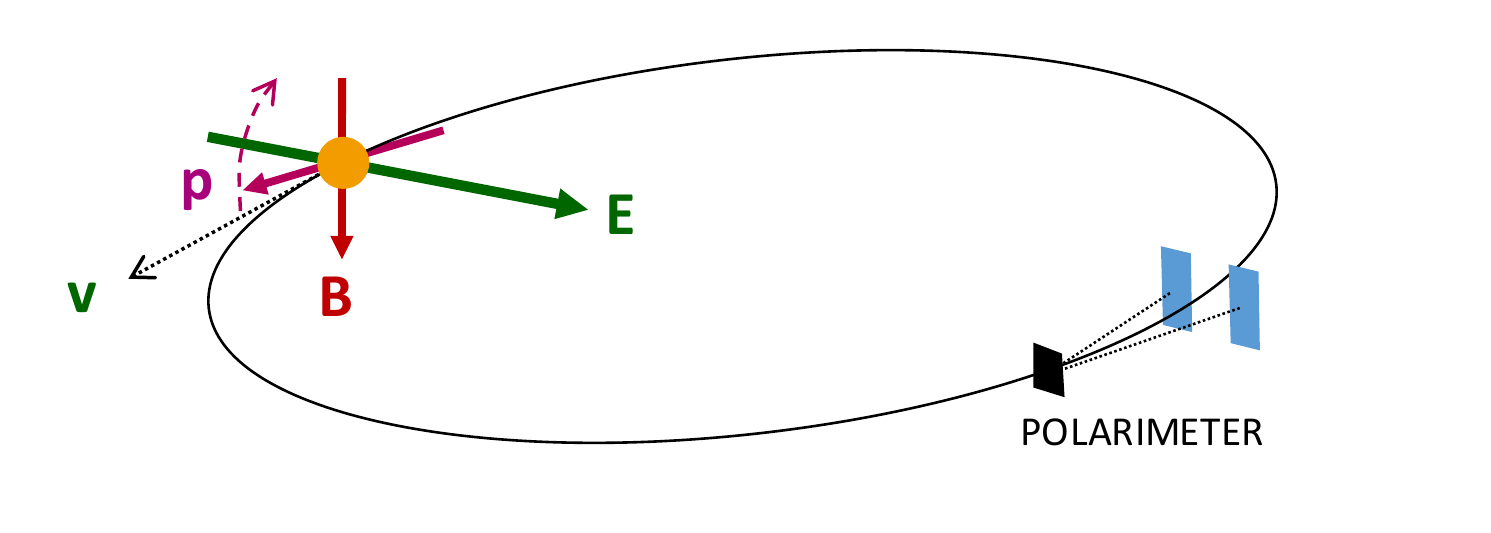}
\caption{Diagram showing a particle
traveling around the storage ring confined by magnetic and electric fields.
The polarization, initially along
the velocity, precesses slowly upward in response to the
radial electric field acting on the EDM. The vertical component of this polarization is
observed through scattering in the polarimeter.}
\end{figure}

The angular frequencies ($\vec\Omega$) in Eq.~(1) are defined with respect to the momentum vector of the particle which itself is changing as the particle travels around its orbit. Because the magnetic moments of all particles carry an anomalous part, the polarization will in general rotate in the plane of the storage ring relative to the beam path. This rotation must be suppressed by making $\vec\Omega_{MDM}=0$, a condition called ``frozen spin''. Otherwise the upward motion of the polarization cannot build up. In a magnetic ring, this condition requires that (since $\vec\beta\cdot\vec B =0$) a radial electric field be added to the ring bending elements with

\begin{equation}
E_r=\frac{GBc\beta\gamma^2}{1-G B\beta^2\gamma^2}\ .
\end{equation}

For particles such as the proton where $G>0$, it is also possible to build an all-electric ring ($\vec B=0$) provided that one can choose $\gamma =\sqrt{1+1/G}$. For the proton, this gives $p=0.7007$~GeV/c. The kinetic energy of $T=232.8$~MeV fortuitously comes at a point where the spin sensitivity of the polarimeter is near its maximum ({\it e.g.} carbon target), creating an advantageous experimental situation.

The statistical error for one single machine cycle is given by~\cite{Ana}
\begin{equation}\label{eq:staterr}
  \sigma_{\mbox{stat}} \approx \frac{2\hbar}{\sqrt{N f} \tau P A E} \, .
\end{equation}
Assuming the parameters given in Table~\ref{tab:staterr}, the statistical error
for one year of running ({\it i.e}. 10000 cycles of 1000\, s length) is
\begin{equation}
    \sigma_{\mbox{stat}}(\mathrm {1 \, year}) = 2.4 \times 10^{-29} \,\text{e} \cdot \mbox{cm} \, .
\end{equation}
The challenge is to suppress the systematic error to the same level.

\begin{table}[htb]
\begin{center}
\caption{Parameters relevant for the statistical error in the proton experiment.
\label{tab:staterr}}
\begin{tabular}{l|l}
\hline
beam intensity &  $N=4 \cdot 10^{10}$ per fill \\
polarization & $P=0.8$   \\
spin coherence time & $\tau = 1000\,$s  \\
electric fields & $E=8\,$MV/m \\
polarimeter analyzing power &  $A = 0.6$ \\
polarimeter efficiency & $f= 0.005$ \\
\hline
\end{tabular}

\end{center}
\end{table}

\begin{wrapfigure}{l}{0.5\textwidth}
\centering{
\includegraphics[width=0.45\textwidth]{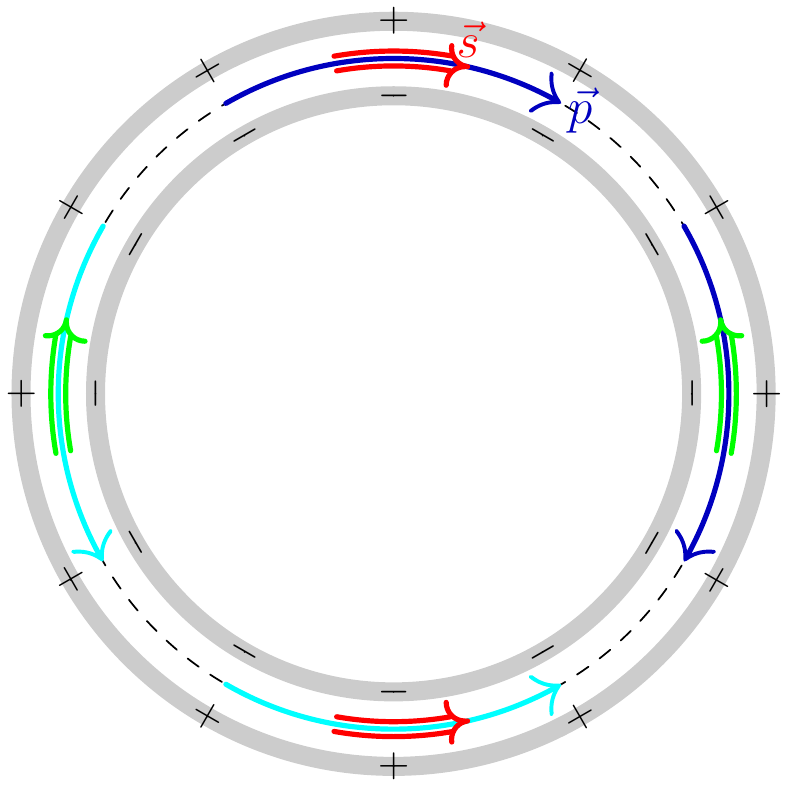}
\caption{Electric storage ring with simultaneously clockwise and counter clockwise circulating beams (dark and light blue arrows), each with two helicity states (green and red arrows for each beam). The gray circles represent electric field plates.  \label{fig:ring_princ}}}
\end{wrapfigure}

A large fraction of the systematic errors in the EDM search may be eliminated by looking at the difference between two experiments run with clockwise (CW) and counter-clockwise (CCW) beams in the ring. One beam represents the time-reverse of the other, and the difference will show only time-odd effects such as the EDM. For the proton, the choice of an all-electric ring makes it possible to have the two beams present together in the ring, an advantage when suppressing systematic effects.
Fig.~2 illustrates two features of the all-electric proton experiment, the counter-rotating beams and the alternating direction of the polarization (along or against the velocity) in separate beam bunches, which is important for geometric error cancellation in the polarimeter.

In general any phenomena other than an EDM generating a vertical component of the spin limits the sensitivity ({\it i.e.} the smallest detectable EDM) of the proposed experiment. Such systematic effects may be caused by unwanted electric fields due to imperfections of the focusing structure (such as the misalignment of components) or by magnetic fields penetrating the magnetic shielding or produced inside the shield by the beam itself, the RF cavity, or gravity. A combination of several such phenomena, or a combination of an average horizontal spin and one of these phenomena, may as well lead to such systematic effects.

In many cases, as for example effects due to gravity, the resulting rotations of the spin into the vertical plane do not mimic an EDM because the observations for the two counter-rotating are not compatible with a time-odd effect. In this case, the contributions from the two counter-rotating beams tend to cancel, provided the forward and reverse polarimeters can be calibrated with sufficient precision. In some cases, as for example magnetic fields from the RF cavity, the resulting spin rotations into the vertical plane can be large.

The most important mechanism dominating systematic effects is an average static radial magnetic field that mimics an EDM signal. For a \mbox{500 m} circumference frozen-spin EDM ring, an average magnetic field of about $10^{-17}$~T generates the same vertical spin precession as the
EDM of $10^{-29}$~e$\cdot$cm the final experiment aims at being able to identify. In order to mitigate systematic effects, the proposed ring will be installed in state-of-the-art magnetic shielding that reduces residual fields to the nT level~\cite{Ana}. The vertical position difference between the two counter-rotating beams that is caused by the remaining radial field will be measured with special pick-ups that must be installed at very regular locations around the circumference to measure the varying radial magnetic field component created by the bunched beam separation. A complete, thorough study of systematic errors in the EDM experiment is very delicate and not yet available. Studies of systematic effects have been carried out and are underway by several teams in the CPEDM collaboration to further improve the understanding of basic phenomena to be taken into account and to estimate the achievable sensitivity. The preliminary conclusion is that the intended sensitivity is very challenging. Meeting this challenge requires that we proceed in a series of stages (see Fig.~3) where each one depends on the knowledge gained from the preceding stage's experience. 

\section{Readiness and Expected Challenges}

The JEDI (J\"ulich Electric Dipole moment Investigations) Collaboration has worked with COSY (COoler SYnchrotron at the Forschungszentrum J\"ulich in Germany) for the last decade to demonstrate the feasibility of critical EDM technologies for the storage ring. Historically, these studies were begun with deuterons, and the switch has not been made yet to protons in order to preserve and build on the deuteron experience. These studies include:
\begin{itemize}[noitemsep,topsep=0pt]
\item[$\bullet$] The beam may be slowly brought to thick ($\sim$~2~cm) target blocks for scattering particles most efficiently into the polarimeter detectors. Favoring elastic scattering events yields the best polarimeter performance~\cite{Bra}.

\item[$\bullet$] After suitable calibration, a comparison of the left-right asymmetries for oppositely polarized beam bunches may be used to lower the polarization systematic error below one part per million~\cite{Bra}.
\item[$\bullet$] Time marking polarimeter events~\cite{Bag} leads to an unfolding of the precession of the in-plane polarization and the measurement of the spin tune ($\nu_S=G\gamma$, polarization revolutions per turn) to a part in $10^{10}$ in a single cycle of 100-s length~\cite{Eve}. The polarimeter signals permit feedback stabilization of the phase~\cite{Hem} of the in-plane precession to better than one part per billion ($10^9$) over the time of the machine store. This is necessary to maintain frozen spin.
\item[$\bullet$] By using bunched beam, electron cooling, and trimming of the ring fields to sextupole order, the polarizataion decoherence with time may be reduced, yielding a lifetime in excess of 1000~s~\cite{Gui}.
Observation of the spin tune variations allows for the measurement of the direction of the invariant polarization axis with a precision of about 1~mrad~\cite{Sal}.
\end{itemize}

With deuterons in the COSY ring at 970~MeV/c 
with non-frozen spin, the polarization precesses in the horizontal plane at 121~kHz relative to the velocity. If there is also an RF Wien filter in the ring that oscillates at the polarization rotation frequency and with its magnetic axis vertical, then some of the EDM signal cancellation may be recovered. This has become the basis for the precursor experiment~(see Fig.~\ref{fig:figure3}, stage 1). Initial running reveals that EDM-like signals do accumulate that arise from systematic perturbations of the deuteron spin as it goes around COSY. These may be compared to the effects of small rotational misalignments of the Wien filter about the beam line and longitudinal polarization changes induced by an away-side solenoid. The EDM signal should resemble a Wien filter rotation, and efforts are underway to understand how well such an effect can be constrained.

Several key technologies are currently under development for the final ring. These include:
\begin{itemize}[noitemsep,topsep=0pt]
\item[$\bullet$] Electrostatic deflector design that requires testing full scale prototypes in a magnetic field with beam to levels of at least
8~MV/m.
\item[$\bullet$] Beam position monitors are needed to operate at a precision of 10~nm for a measurement time of 1000~s.
 \item[$\bullet$] The ring must be shielded to provide isolation from systematic radial magnetic fields to the nT level~\cite{Ana}.
\end{itemize}

Spin tracking calculations are needed to verify the level of precision needed in the ring construction and the handling of systematic errors.
For a detailed study during beam storage and buildup of the EDM signal, one needs to track a large sample of particles for billions of turns. 
The COSY-Infinity~\cite{cosyi} and Bmad~\cite{bmad} simulation programs are utilized for this purpose.
Given the complexity of the tasks, particle and spin tracking programs 
have been benchmarked and simulation results compared to beam and spin experiments at COSY 
to ensure the required accuracy of the results.

Finally, a strategy will be needed to verify any signal produced by the experiment after the CW-CCW subtraction through a series of critical tests and independent analyses.

When constructed, the proton EDM experiment will be the largest electrostatic ring ever built. It will have unique features, such as counter-rotating beams and strenuous alignment and stability requirements. It may also require stochastic cooling and weak magnetic focusing consistent with dual beam operation. Intense discussions within the CPEDM collaboration have concluded that the final ring cannot be designed and built in one step; instead, a smaller-scale prototype ring (see Fig.~3, stage~2) must be constructed to confirm and refine these critical features:
\begin{itemize}[noitemsep,topsep=0pt]
\item[$\bullet$] The ring stores high beam intensities for a sufficiently
long time.
\item[$\bullet$] Beam injection must allow for multiple polarization states (longitudinal fore and aft, sideways for polarization coherence monitoring) in both CW and CCW beams.
\item[$\bullet$] The ring must circulate CW and CCW beams simultaneously, both horizontally polarized.
\item[$\bullet$] The ring must support frozen spin.
\item[$\bullet$] Magnetic shielding must operate to reduce the ambient magnetic fields (esp.\ radial) to suitable levels while allowing full operation of the ring high voltage, vacuum, monitoring, and control.
\item[$\bullet$] Polarimeter measurements must be made for both CW and CCW beams using the same target. A second polarimeter is needed with independent beam extraction onto its target.
\item[$\bullet$] Beam cooling (electron-cooling before injection, or stochastic) is required to reduce the beam phase space.
\end{itemize}

\section{CPEDM Strategy}

\begin{figure}[htb]
\centering
\includegraphics[scale=0.54]{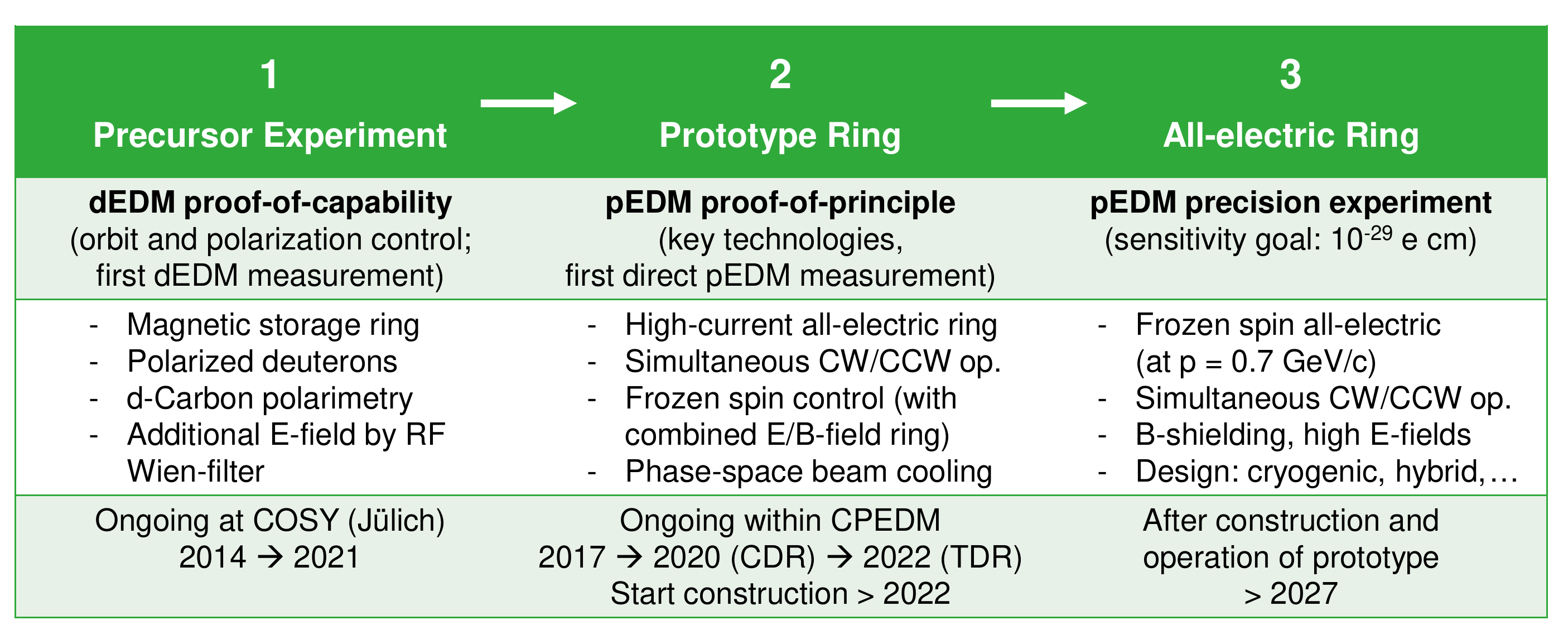}
\caption{\label{fig:figure3}Summary of the important features of the proposed stages in the storage ring EDM strategy.}
\end{figure}

As emphasized above this challenging project needs to proceed in stages
that are also outlined in Fig.~\ref{fig:figure3}:

\begin{enumerate}[noitemsep,topsep=0pt]
\item COSY will continue to be used as long as possible for the continuation of critical R\&D associated with the final experiment design. An important requirement is to test as many of the results as possible with protons where the larger anomalous magnetic moment leads to more rapid spin manipulation speeds.
\item The precursor experiment will be completed and analyzed. Some data will be taken with an improved version of the Wien filter with better electric and magnetic field matching.
\item The next stage is to design, fund, and build a prototype ring (discussed in detail below) to address critical questions concerning the features of the EDM ring design. At 30~MeV, the ring with only an electric field can store counter-rotating beams, but they are not frozen spin. At 45~MeV with an additional magnetic field, the frozen spin condition can be met. But the magnetic fields also prevent the CW and CCW beams from being stored at the same time. Even so, an EDM experiment may be done with these two beams used on alternating fills.
\item Following step 3, the focus will be to create the final ring design, then fund and construct it.
\item Once the ring is ready, the longer term activity will be to commission and operate the final ring, improving it with new versions as the systematic errors and other experimental issues are understood and improved.
\end{enumerate}

Future scientific goals may include
conversion of the ring to crossed electric and magnetic field operation so that other species besides the proton could be examined for the presence of an EDM. Analysis
of the data may be made for signs of axions using a frequency decomposition and investigation of counter-rotating beams with different species used in novel EDM comparisons.

The prototype ring and the CPEDM stages are host-independent.
If the prototype is built at COSY, it would take advantage of the existing facility for the production of polarized proton (and deuteron) beams, beam bunching, and spin manipulation. COSY itself could be used for producing electron-cooled beams. It may also be built at another site ({\it e.g.}, CERN) provided that a comparable beam preparation infrastructure is made available. In either case, the lattice design will mimic that of the high-precision ring in order to test as many features as possible on a smaller scale. 

\section{Details of the Prototype EDM ring}

The prototype (PT) ring will be small (circumference of 100~m) and 
operate in two modes (see stage 3 above and Table 2). The ring will be as inexpensive as possible, consistent with being capable of achieving its goals. The first mode would operate with all-electric
bending (at 30~MeV), a demonstration that such a concept works and may be
used to demonstrate feasibility of the ring with simultaneous
counter-rotation beams. The second would extend the operating
range to 45 MeV with the addition of magnetic bending (air core).
With this combination, frozen spins could be demonstrated for a
proton beam, other spin manipulation  
tools developed, and a reduced-precision proton EDM value measured. 
Alternating fills in counter-rotating directions would allow cancellation of the average radial magnetic field $\langle B_r\rangle$ 
that is the leading cause of systematic error (though with a large 
systematic error associated with the needed magnetic field reversal).

This section describes a starting-point lattice in terms of geometry, type and strength of the elements. The ring is square with 8~m long straight sections. The basic
beam parameters are given in Table~2 and a rough tabulation of costs is given in the Addendum.

\begin{table}[h]
\centering
\caption{Basic beam parameters for the PT ring}
\label{tab:table1}
\begin{tabular}{l|c|c|c}
& $E$ only & $E\times B$ & unit \\
\hline
kinetic energy & 30 & 45 & MeV \\
\hline
$\beta =v/c$ & 0.247 & 0.299 & \\
momentum & 239 & 294 & MeV/c \\
magnetic rigidity $B\rho$ & 0.798 & 0.981 & T$\cdot$m \\
electric rigidity $E\rho$ & 59.071 & 87.941 & MV \\
$\gamma$ (kinetic) & 1.032 & 1.048 & \\
emittance $\epsilon_x =\epsilon_y$ & 1.0 & 1.0 & mm$\cdot$mrad \\
acceptance $a_x=a_y$ & 1.0 & 10.0 & mm$\cdot$mrad \\
\hline
\end{tabular}
\end{table}


\subsection{PT requirements and goals}

The foremost goal is to demonstrate the ability to store enough protons ($10^{10}$) to be able to perform proton EDM
measurements in an electric storage ring, recognizing that some superimposed magnetic bending
is likely to be necessary to meet this goal.

Since ultimate EDM precision will require simultaneously counter-circulating beams a prototype ring has to
demonstrate the ability to store and control simultaneously two such beams.

Cost-saving measures in the prototype, such as room temperature operation, minimal magnetic shielding, and 
avoidance of obsessively tight manufacturing and field-shape matching tolerances, are expected to limit the
precision of any prototype ring EDM measurement. Nonetheless, data for reliable cost estimation and extrapolation
of the systematic error evaluation to the full scale ring has to be obtained.




\subsection{PT ring design}

The basic parameters for the PT ring are given in Tables~3 and 4. The bending, for example
for 45~MeV protons, is done by eight $45^{\circ}$ electric/magnetic bending elements. The acceptance of the ring
is to be 10~mm$\cdot$mrad for $10^{10}$ particles. The lattice has fourfold symmetry, as shown in Fig.~4. The lattice is
designed to allow a variable tune between 1.0 and 2.0 in the radial plane and between 1.6 and 0.1 respectively in the vertical
plane.

\begin{table}[!htb]
\begin{minipage}[t]{.5\textwidth}
\centering
    \caption{\label{tbl:Table2}Geometry}
    \begin{tabular}[t]{l|l|l}
    & & units \\
    \hline
    \# B-E deflectors        & 8       &     \\
    \# arc D quads           & 4       &     \\
    \# arc F quads           & 8       &     \\
    quad length              & 0.400   &  m  \\
    straight length          & 8.000   &  m  \\
    bending radius           & 8.861   &  m  \\
    electric plate length    & 6.959   &  m  \\
    arc length ($45^{\circ})$ & 15.718  &  m  \\
    circumference total      & 100.473 &  m  \\
    \hline
    \end{tabular}
\end{minipage}
\begin{minipage}[t]{.4\textwidth}
\centering
    \caption{\label{tbl:Table3}Bend elements}
    \begin{tabular}[t]{l|l|l}
    & & units \\
    \hline
    {\bf Electric}            &              &          \\
    \hline
    gap between plates        & 60           & mm       \\
    plate length              & 6.959        & m        \\
    total bending length        & 55.673       & m        \\
    total straight length    & 44.800       & m        \\
    bend angle per unit       & $45^{\circ}$  &  m       \\
    \hline
    {\bf Magnetic}            &              &          \\
    \hline
    magnetic field            & 0.040        &  T       \\
    current density           & 5.000        & A/mm$^2$ \\
    windings/element          & 60           &          \\
    \hline
    \end{tabular}
\end{minipage}
\end{table}
%

\newpage
\begin{wrapfigure}{l}{0.5\textwidth}
\centering
\includegraphics[scale=0.48
,viewport=100 350 600 775,clip]{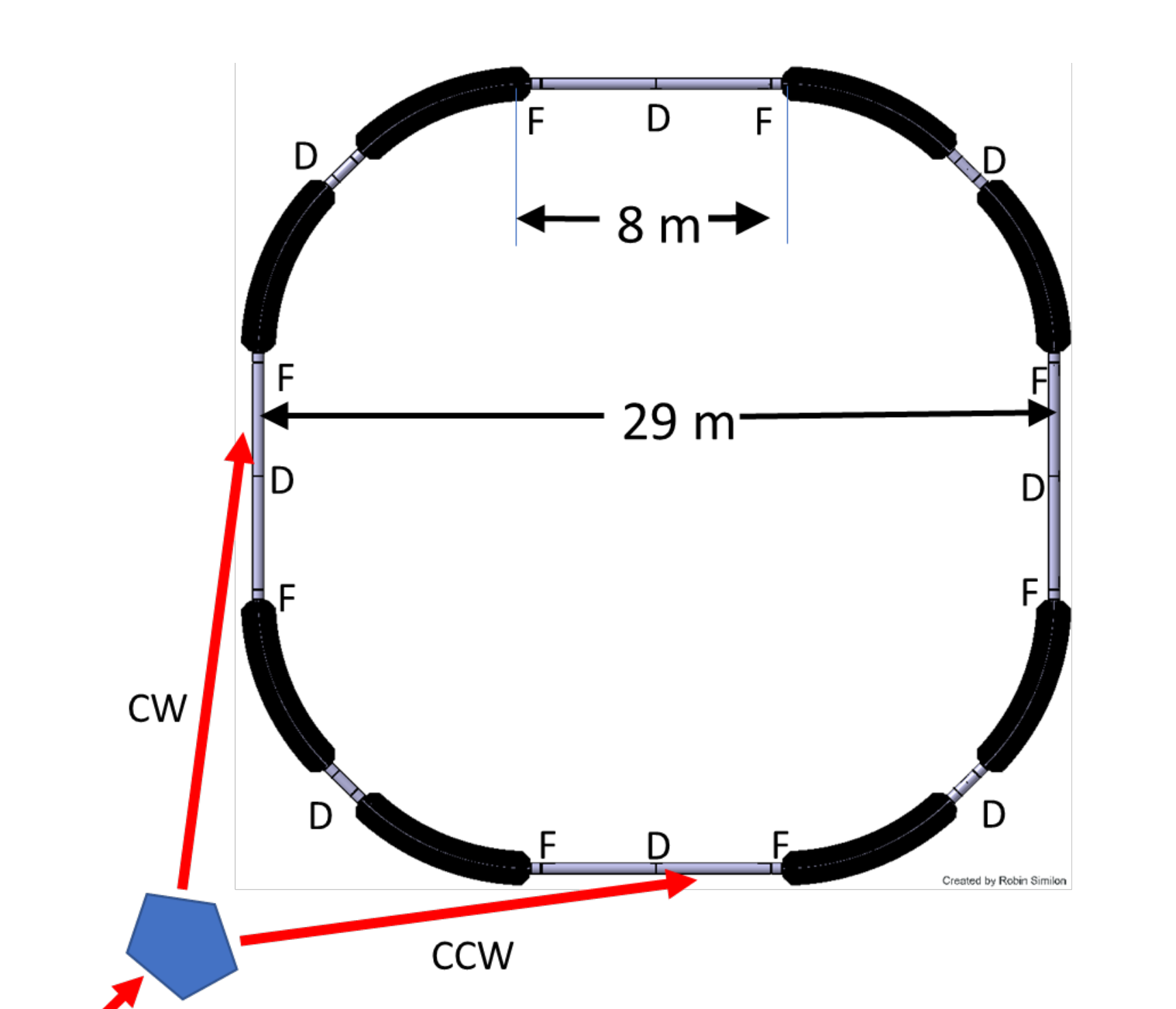}
\caption{The basic layout of the prototype
ring, consisting of 8 dual, superimposed
electric and magnetic bends; 3 families of
quadrupoles (Focusing, Defocusing, and Straight-section); and four 8-m long
straight sections. The total circumference is
about 100 m. Injection lines for injecting counter-circulating
beam are represented just as stubs. Costs given in the Addendum 
are restricted to just the PT ring, which is truly site-independent. 
The possibly greater infrastructure costs associated with
producing appropriately polarized beams are neither given
nor site-independent.}
\end{wrapfigure}

\noindent {\bf The injector:} Injection into the PT ring will closely resemble injection into a nominal all-electric ring.
In particular there will be an even number of bunches in each beam, with alternating sign polarizations, whether
in single beam or counter-circulating beam operation. The injector for the PT ring could be the electron-cooled beam from COSY or make use of equipment at CERN. The beams will be
protons in the 30 to 45 MeV range, in a cooled phase space of 1~$\pi$mm$\cdot$mr, with the beams bunched into 2, 4, 6
or 8 bunches to be fed into counter-circulating beams in the PT ring.

Injection into the prototype ring will be done using switching magnets distributing the beams into clockwise
(CW) or counter clockwise (CCW) direction as sketched in Fig.~4. All beam bunches are transferred with
vertical polarization, either up or down.

\vskip0.05in
\noindent {\bf Electric bends:} The electrostatic deflectors consist of two cylindrically-shaped parallel metal plates
with equal potential and opposite sign. With the zero voltage contour of electric potential defined to be the center
line of the deflector, the ideal orbit of the design particle stays on the center line. The electrical potential vanishes
on the center line of the bends, as well as in drift sections well outside the bends. So the electric potential vanishes
everywhere on the ideal particle orbit. With the electric potential seen by the ideal particle continuous at the
entrance and exit of the deflector, its total momentum is constant everywhere (even through the RF cavity).

The designed ring lattice requires electric gradients in the range from 5 to 10~MV/m. This is more than the standard
values for most accelerator deflectors separated by a few centimeters. Assuming 60~mm distance between the
plates, to achieve such high electric fields we have to use high voltage power supplies. At present, two 200~kV
power converters have been ordered for testing deflector prototypes. The field emission, field breakdown,
dark current, electrode surface and conditioning will be studied using two flat electrostatic deflector plates,
mounted on the movable support with the possibility of changing the separation from 20 to 120~mm. The residual
ripple of power converters is expected to be in the order of $\pm 10^{-5}$ at a maximum of 200~kV. This will lead to particle
displacement on the order of millimeters. A smaller ripple or stability control of the system will be a dedicated
task for investigations planned at the test ring facility.
\vskip0.05in
\noindent {\bf Magnetic bends:} The experiments require periodic reversals of the magnetic bending field to use symmetry to
suppress systematic deviations. The reversal of the magnetic field should be done with best possible reproducibility.
This is why the magnetic field production will iron-free
(see Fig.~5).

\begin{figure}[htb]
\centering
\includegraphics[scale=0.63]{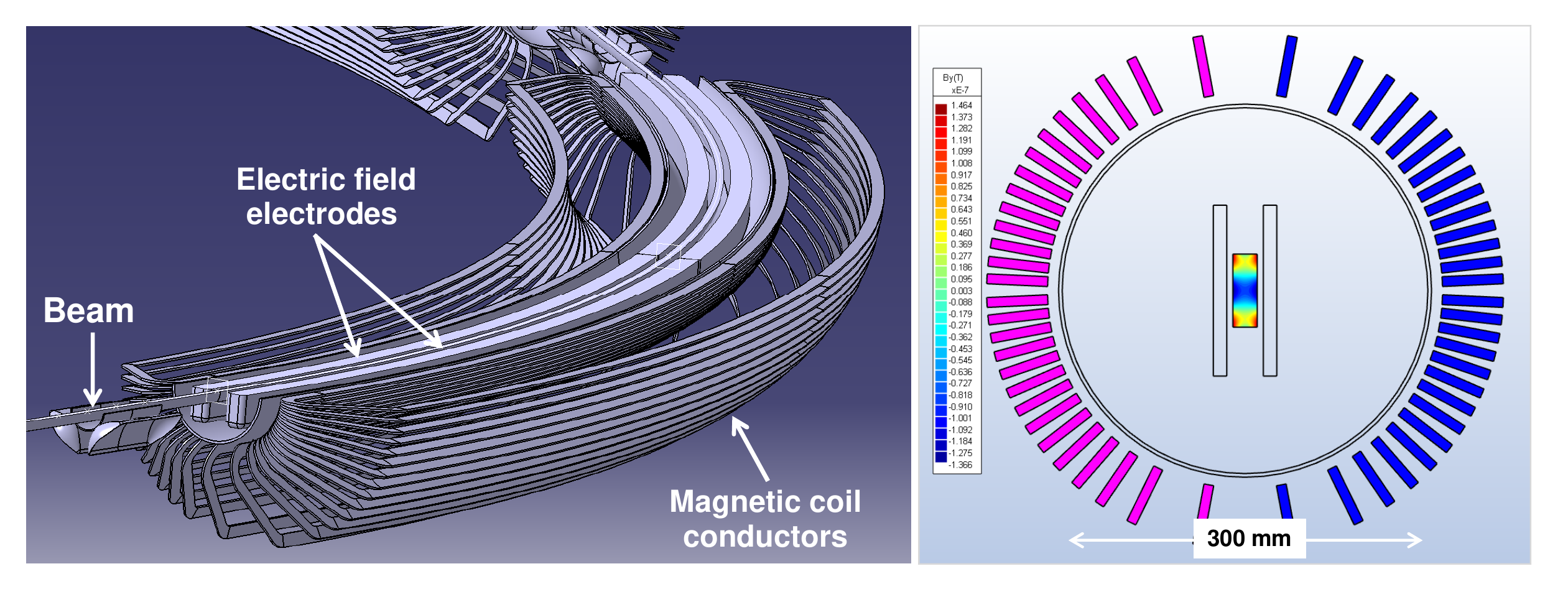}
\caption{Shown on the left is a cutaway drawing of the PT ring in the $E\times B$ version. A side view of the lower half of a $ 45^{\circ}$
bend element is shown. The electrodes have a gap of 60~mm. The magnetic coil conductors (single, $4\times 4$~mm$^2$ copper bars) produce a highly uniform ``cosine-theta'' dipole field.  Shown on the right is a transverse
section showing an end view of the (inner legs of the) magnet coil, as well as a field map of the good magnetic field region.}
\end{figure}


\noindent {\bf Other components:} All quadrupoles will be electrostatic. Their design will follow the principles of the Heidelberg
CSR ring~\cite{CSR}. Both DC and AC Wien filters and solenoids will be required for spin control. The RF cavity design
is under study.


The requirement for the vacuum is mainly given by the minimum beam lifetime requirement of about 1000~s.
The emittance growth in the ring caused by multiple scattering from the residual gas is\break 0.005~mm$\cdot$mrad/s. At
$10^{-12}$~Torr vacuum, the emittance at the beginning, assumed to be 1~mm$\cdot$mrad, will have increased to 5~mm$\cdot$mrad
within 1000~s, assuming a nitrogen (N$_2$) partial pressure. This is about the cooling rate expected for stochastic
cooling. (One notes in passing that stochastic cooling becomes impractical for very low tunes.) For
such an ultra-high vacuum only cryogenic or NEG pumping systems can be used. Bake-out must be foreseen
for either cryogenic or NEG systems. 


The choice of NEG requires a beam pipe with a diameter of 300~mm over the full circumference of 100~m. This
can easily be plated with the NEG material. We will then have an active area of $\approx 120$~m$^2$ for the whole ring. The
roughing speed will be about 5000~liter/s per meter of length of vacuum pipe.


There are beam position monitors located around the ring. A BPM is placed at the entrance and the
exit of each bending unit. One BPM will be placed additionally in close connection to the quadrupoles in the
straight sections. A new type of BPM, of Rogowski coil design~\cite{Tri}, has been developed at the IKP of the FZ-J\"ulich.
These pick-ups are presently in a development stage. The position resolution is measured to be 10~$\mu$m over an area with a diameter of about 90~mm. These BPMs require only a short beam insertion length of 60~mm and an offset-bias free
response to counter-circulating beams.

\section{All-electric storage ring}

This document describes the vision of CPEDM culminating in the design, construction, and operation of a dedicated, high-precision storage ring for protons. Operating at the all-electric, frozen-spin momentum of 0.7~GeV/c, the signals from counter-rotating beams aim to measure the proton electric dipole moment with a sensitivity of $10^{-29}$~e$\cdot$cm.  The major challenge
is the handling of all systematic errors to obtain an overall sensitivity of a similar size. The main source of systematic uncertainties will be due to any unknown or unidentified radial magnetic field acting through the much larger magnetic dipole moment and leading to a false EDM signal. The level at which this can be mitigated remains to be determined.

Invaluable results and experiences are expected from the intermediate step, the construction of a smaller, prototype ring. The attempts to examine the control of counter-rotating beams and study directly the conditions for frozen spin will have a huge impact on the detailed outline of the high-precision ring design.

The concept of an all-electric storage ring with extremely well-fabricated and aligned elements running two longitudinally polarized proton beams in opposite directions in the absence of significant magnetic fields serves as the current starting point. There are new ideas under development that offer the prospect of further mitigation of the systematic issues:
\begin{itemize}
\item[$\bullet$] A hybrid electric/magnetic ring with magnetic focusing (in addition to electric deflector contributions) will change the electro-magnetic environment in significant ways. Even in the presence of uncontrolled radial magnetic fields, this geometry offers at least one point at which the magnetic field vanishes. Beam-based alignment techniques will tend to find these points and place the beam there. This substantially relaxes the requirement that radial magnetic fields be made to nearly vanish. The magnetic focusing, however, does not produce 
counter-rotating beams with the same phase space profile. So periodic reversal of the magnetic focusing would be required to provide a set of signals that must be averaged to obtain an EDM value.
\item[$\bullet$] It is possible to find pairs of unlike polarized beams for which the same superimposed electric and magnetic bending yields a frozen spin condition for both (e.g. protons and $^3$He). Since the two beams would not have the same revolution frequency, to circulate simultaneously they would run with appropriately different RF harmonic numbers. Though not yielding either EDM value directly, the resulting EDM difference will be independent of the (otherwise dominant) radial magnetic field systematic error. Any EDM signal differences would be interpreted as the presence of an EDM on at least one of the two beams. 
\end{itemize}
Work on these concepts can proceed using the prototype ring with the possibility of yielding new physics results.

The storage ring EDM feasibility studies made so far show encouraging results. Handling systematic error signals is the main challenge. The path to addressing this lies through the construction and operation of a small-scale prototype ring from which will come the design for the high-precision ring with the best sensitivity to new physics.



\end{document}